\documentclass[amsmath,amssymb,superscriptaddress,prl,reprint]{revtex4-1}

\usepackage[utf8]{inputenc}
\usepackage{graphicx}
\usepackage{xcolor}
\usepackage{hyperref}
\usepackage{notes2bib}
\usepackage{siunitx}
\DeclareSIUnit\Molar{\textsc{m}}

\newcommand{\circled}[1]{\raisebox{.5pt}{\textcircled{\raisebox{-.9pt} {#1}}}}

\begin{document}
\

\title{Average negative heat in a non-Markovian bath}
\author{F\'elix Ginot}
\email{felix.ginot(at)uni-konstanz.de}
\affiliation{Fachbereich Physik, Universit\"{a}t Konstanz, 78457 Konstanz, Germany}
\author{Clemens Bechinger}
\affiliation{Fachbereich Physik, Universit\"{a}t Konstanz, 78457 Konstanz, Germany}

\begin{abstract}
We experimentally study the motion of a colloidal particle, translated back and forth within a viscoelastic, i.e. non-Markovian bath. 
The particle starts in equilibrium before the forward motion, but only partially relaxes at the turning point. 
During the backward motion, we measure a systematic (negative) heat flow from the bath to the particle. 
Our observations are in good agreement with a simple model that describes the time-delayed response of the fluid. 
We expect our results to be important for the realization and optimization of novel types of micro-engines in non-Markovian surroundings.
\end{abstract}

\maketitle


Stochastic thermodynamics provides a powerful framework to describe the behaviour of small systems where fluctuations become crucial ~\cite{sekimoto1998langevin,seifert2012stochastic}. 
Within this approach, the concepts of classical, i.e., macroscopic thermodynamics can be transferred to microscopic length scales which permits the definition of small-scale equivalents of entropy, work, heat or internal energy ~\cite{crooks1999entropy,seifert2005entropy,baiesi2009fluctuations, jarzynski2011equalities}. 
Opposed to macroscopic systems, however, such quantities are not given by sharp values but by distributions with finite width which is an immediate consequence of thermal fluctuations which govern the behavior of tiny systems. 

An immediate consequence of these fluctuations is the temporal "violation" of the second law of thermodynamics. 
Experimental studies~\cite{collin2005verification,bustamante2005nonequilibrium,berut2012experimental} of colloidal particles driven through water revealed clear evidence that entropy can be consumed rather than generated on short time scales~\cite{wang2002experimental,evans2002fluctuation}. 
Nevertheless, because the viscous friction of a dragged particle is entirely dissipated within the surrounding heat bath, the average entropy and heat production in such experiments always remains positive.
This, however, is only valid when the relaxation time of the bath is much shorter than that of the particle, i.e. when the bath remains in equilibrium even while the particle is driven~\cite{dexter1972mechanical}.
In the case of a viscoelastic bath with long relaxation time, such conditions are no longer fulfilled~\cite{larson1999structure,liu2006microrheology,gomez2014probing,khan2019optical,ginot2022recoil}.
As such environments constitute the natural surrounding for molecular motors~\cite{schliwa2003molecular}, bacteria~\cite{peterson2015viscoelasticity} and motile cells~\cite{thurston1972viscoelasticity}, this breaking has important consequences on the heat production for driven systems at small scales.

In this work, we experimentally demonstrate that the heat of a colloidal particle driven through a viscoelastic fluid can be negative, even when looking at averages. 
Because the relaxation time $\tau_R$ of the fluid is several seconds, it cannot immediately dissipate the energy of a driven particle. 
As a consequence, a finite amount of the transferred energy is restored back to the particle, which leads to an averaged negative heat (ANH) production which is limited by the amount of work initially spent on the particle. 
Our results are in excellent agreement with a micro-mechanical~\cite{darabi2023stochastic} model, which is expected to be also applicable to other types of non-Markovian baths. We expect our findings to be particularly relevant for the design and optimization of thermodynamic machines in non-Markovian environments.


We perform experiments using silica particles with diameter $\SI{2.73}{\micro\m}$ within a \SI{100}{\micro\m} thick capillary.
The particles are suspended in an \SI{8}{\milli\Molar} aqueous solution of cetylpyridinium chloride monohydrate (CPyCl) and sodium salicylate (NaSal).
We kept the sample cell at \SI{25}{\celsius} where the fluid forms an entangled network of giant worm-like micelles leading to a viscoelastic, i.e. time-delayed behavior~\cite{cates1990statics}. 
At the above conditions, the largest relaxation time of a \emph{free} particle (in absence of a trap) has been determined to be $\tau_R\approx \SI{2}{\s}$ \cite{ginot2022recoil}. 
Note, that the relaxation time is considerably longer in presence of a trap. 
We confined the particle using a highly focused \SI{1064}{\nano \m} laser beam, leading to a harmonic optical potential $V(x,\lambda) = \frac{1}{2}\kappa (x-\lambda)^2$ with stiffness $\kappa=\SI{2.0}{\micro\N\per\m}$.
The trap position $\lambda(t)$ was controlled using a piezo-actuated stage, and kept far from the surface of the capillary.
We recorded video pictures with a frame rate of \SI{100}{\hertz}, and obtained particle trajectories $X(t)$ with an accuracy of $\pm \SI{6}{\nano\m}$ using a custom Matlab tracking routine~\cite{crocker1996methods}.
More information regarding sample preparation and the setup, is available in Supplemental Material (SM).

\begin{figure}
    \centering
    \includegraphics[width=8.6cm]{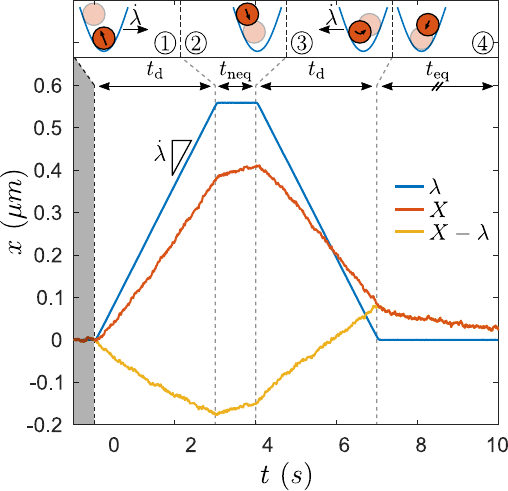}
    \caption{Time dependent average position of the trap $\lambda(t)$ (blue), and of the particle $X(t)$ (red) during a typical experimental protocol. \circled{1} Starting from equilibrium, the trap moves at constant speed $\dot \lambda$ for a time $t_\text{d}$. \circled{2} The particle partially relaxes in the trap for a time $t_\text{neq}=\SI{1}{\s}$. \circled{3} The trap is put back to its initial position at speed $-\dot \lambda$. \circled{4} Finally, the system relaxes to equilibrium for a time $t_\text{eq}$ (not fully shown here). The orange line corresponds to the particle's position relative to the trap $X-\lambda$, which is proportional to the optical force exerted on the particle.
    }
    \label{fig:F1}
\end{figure}

 Figure \ref{fig:F1} shows the typical driving cycle protocol being applied in our experiments, which consists of four steps.
\circled{1} Starting from a fully equilibrated state, we first move the trap to the right at constant speed $\dot \lambda=\SI{0.2}{\micro\m\per\s}$ during time $t_\text{d} = \SI{3}{\s}$. 
\circled{2} The trap motion is stopped for the time $t_\text{neq}$ during which the particle relaxes within the trap. 
However, since $t_\text{neq}$ is smaller than the time required for equilibration, this relaxation process is not complete, and the trap exerts a non-zero force on the particle (see orange line). 
\circled{3} Afterward, the protocol is reversed, i.e. the trap moves left, back to its original position with opposite velocity $\dot \lambda=\SI{-0.2}{\micro\m\per\s}$. 
\circled{4} Finally, we let the system fully equilibrate during time $t_\text{eq}=\SI{50}{\s}$ before the next cycle starts.
Due to $t_\text{eq} >> t_\text{neq}$, the forward-backward protocol is asymmetric which can be seen on the particle trajectory (red line). 
To yield sufficient statistics, the cycle is repeated about 100 times. 

From the trap and particle's trajectory, we calculate the work $W$ and heat $Q$ associated with the particle with potential energy $U[X(t)] = \frac{1}{2}\kappa (X-\lambda)^2$.
The accumulated values within the time interval $[0,t]$  are given by~\cite{seifert2012stochastic,speck2007jarzynski} 
\begin{equation}
	W[X(t)] =  \int_0^t \frac{\partial V}{\partial \lambda}\dot\lambda\text{d}t = - \int_0^t \kappa(X-\lambda)\dot\lambda\text{d}t \label{dw eq}
\end{equation}
and 
\begin{equation}
	Q[X(t)] = \int_0^t \frac{\partial V}{\partial X}\dot X\text{d}t = \int_0^t \kappa(X-\lambda)\dot X\text{d}t\label{dq eq}.
\end{equation}


\begin{figure}
    \centering
    \includegraphics[width=8.6cm]{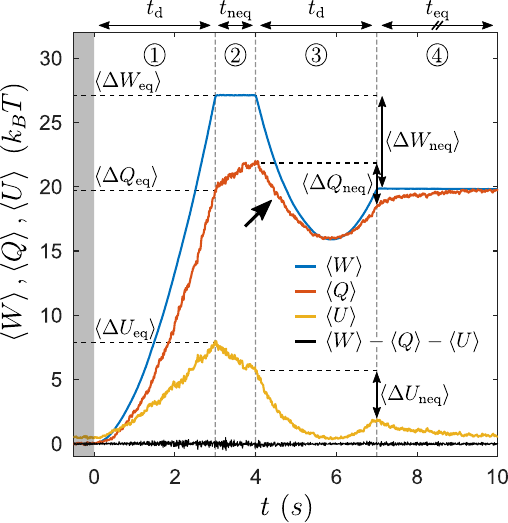}
    \caption{Time dependent average work $\left<W(t)\right>$ (blue), heat $\left<Q(t)\right>$ (red) and internal energy $\left<U(t)\right>$ (orange) for an experiment with $t_\text{neq}=\SI{1}{\s}$. Black line corresponds to the difference between work and the sum of heat plus internal energy, which adds up to zero in accordance to the first law of thermodynamics.}
    \label{fig:F2}
\end{figure}

Figure \ref{fig:F2} shows the average time-dependent values of work $\left<W(t)\right>$ (blue), heat $\left<Q(t)\right>$ (red) and internal energy $\left<U(t)\right>$ (orange) for $t_\text{neq} = \SI{1}{\s}$. 
The system starts at equilibrium and $\left<U(t=0)\right> = \frac{1}{2}k_BT$. 
During \circled{1} the trap moves with constant speed, the viscous force of the fluid causes the particle to slightly lag behind the trap center, and $\left<U(t)\right>$ increases. 
Since the corresponding optical force acting on the particle is then pointing opposite to the trap motion, $\left<W(t)\right>$ also increases (Eq.\ref{dw eq}). 
Because $\left<W(t)\right> > \left<U(t)\right>$, the heat $\left<Q(t)\right>$ is positive (energy is flowing from the system towards the bath) in agreement with the first law $W - Q - U = 0$ (see black line). 
In the following the changes of work, heat, and internal energy during step \circled{1} will be denoted as $\Delta W_\text{eq}$, $\Delta Q_\text{eq}$, and $\Delta U_\text{eq}$ respectively.
During \circled{2} the trap is at rest, $\left<W(t)\right>$ remains constant because $\dot \lambda = 0$ (Eq.~\ref{dw eq}). 
Since the particle is partially relaxing towards the trap center, this leads to an increase of $\left<Q(t)\right>$ at the expense of $\left<U(t)\right>$.
During \circled{3} the trap moves back to its initial position, all quantities first decrease and then increase again at about $t \approx \SI{6}{\s}$.
For $\left<U(t)\right>$ and $\left<W(t)\right>$, this behavior is easily understood by considering that the particle, which is only partially relaxed at the beginning of step \circled{3}, is initially sitting in the front of the advancing trap center. 
After $t \approx \SI{6}{\s}$ the trap has passed the particle position, and both values increase, similar to \circled{1}. 
However, even with the presence of viscous friction during \circled{3} ($\dot \lambda \neq 0$), the heat $\left<Q(t)\right>$ is not monotonically increasing. 
Instead, similar to $\left<W(t)\right>$ and $\left<U(t)\right>$, it first decreases (see arrow in Fig.~\ref{fig:F2}) and only later increases again. 
In the following, we refer to this anomalous transfer of heat, from the bath to the particle, as average negative heat (ANH). 
Changes of work, heat, and internal energy during step \circled{3} will be denoted as $\Delta W_\text{neq}$, $\Delta Q_\text{neq}$, and $\Delta U_\text{neq}$ respectively.
The appearance of ANH which is here reported for a viscoelastic fluid, is absent in viscous, i.e. memory-free baths, where negative heat events are only sporadically observed at the level of single trajectories but not for their averages (see SM). 
As will be shown below, an ANH results from the time-delayed response of a viscoelastic fluid, which prevents heat to become immediately dissipated in the bath. 
As a consequence, heat can be partially recovered from the bath to perform work on the particle.


To rationalize the above experimental findings, we perform numerical simulations using a minimal model. 
The interaction between the colloidal particle and the viscoelastic fluid is introduced using a (fictitious) bath particle which is connected to the colloid via a harmonic spring. 
Accordingly, the extension of the spring is associated with the storage of elastic energy within the bath. 
This microscopic equivalent of the Maxwell model has been previously used to describe the behavior of the viscoelastic solution~\cite{ginot2022barrier,ginot2022recoil}. 
Considering the presence of an external potential $V(x)$, the positions of the colloidal ($X$) and bath ($X_\text{b}$) particles are described by the following Langevin equations:
\begin{eqnarray}
	\gamma\dot{X} &=& -\kappa_\text{b} (X - X_\text{b}) -\nabla V  + \xi(t)  \label{LangevinEquationTracer} \\
	\gamma_{\rm b}\dot{X}_\text{b} &=& -\kappa_\text{b} (X_\text{b} - X)  + \xi_{\text{b}}(t) \label{LangevinEquationBath}
\end{eqnarray}
where $\gamma$ and $\gamma_\text{b}$ correspond to the friction of the colloidal and the bath particle, respectively, and $\kappa_\text{b}$ is the coupling strength. $\xi$ and $\xi_\text{b}$ are delta correlated random forces with zero mean.
Notably, the two coupled Markovian equations correspond to a single non-Markovian generalized Langevin equation with an exponentially decaying memory kernel~\cite{medina1987generalized,mason1995optical}. 
The parameters $\gamma$, $\gamma_\text{b}$, and $\kappa_\text{b}$ have been obtained by comparison with the experimentally measured work, heat and internal energy as shown in Fig.~\ref{fig:F2} (see SM).

Within the above model, one can immediately understand the presence of an ANH. 
When the colloidal particle is driven by the optical trap \circled{1}, the spring to the bath particle becomes extended, and the associated elastic energy increases.
This is a clear example of strong coupling, where driving the system out of equilibrium also sets the bath out of equilibrium.
When the motion of the trap stops \circled{2}, both the colloidal and the bath particles begin to relax. 
However, because $t_\text{neq}$ is too small to reach equilibrium, a finite amount of elastic energy remains trapped in the spring. 
Consequently, upon reversing the driving force on the colloid \circled{3}, the previously extended spring is compressed, which transfers elastic energy back to the colloidal particle. 
When measuring the stochastic heat, this flow of energy from the bath to the colloid translates into an ANH. 
Naturally, the recovered energy cannot exceed the energy previously stored in the spring.
Thus, during a full cycle heat and entropy systematically increases, in agreement with the second law. 


\begin{figure}
    \centering
    \includegraphics[width=8.6cm]{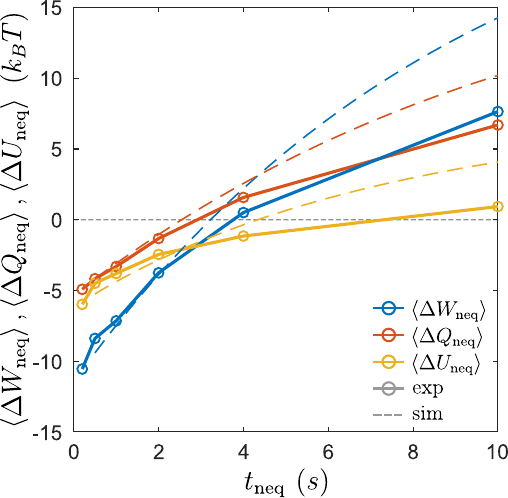}
    \caption{$\left< \Delta W_\text{neq} \right>$ (blue), $\left< \Delta Q_\text{neq} \right>$ (red), and $\left< \Delta U_\text{neq} \right>$ (orange) for increasing values of $t_\text{neq}$. All curves start negative and increase monotonically, albeit at a different rate. For short $t_\text{neq}$ heat and internal energy are converted into work, while the opposite happens for longer $t_\text{neq}$. Simulations (dashed-lines) agree well with experiments (symbols), although deviations are observed for larger values of $t_\text{neq}$.
    }
    \label{fig:F3}
\end{figure}

Another consequence of our model is, that the stored elastic energy determines the magnitude of the ANH. 
For an experimental check, we have varied $t_\text{neq}$ which controls the amount of elastic energy left in the bath at the begin of \circled{3}. 
Figure \ref{fig:F3} shows the experimentally measured $\left<\Delta Q_\text{neq}\right>$ during \circled{2} as a function of $t_\text{neq}$. 
The averaged heat $\left<\Delta Q_\text{neq}\right>$ increases with $t_\text{neq}$ and goes from negative to positive values. 
The latter confirms that ANH is a non-equilibrium feature being absent in fully equilibrated systems. 
In addition we also plotted the average work $\left<\Delta W_\text{neq}\right>$ and internal energy $\left<\Delta U_\text{neq}\right>$ exchanged during \circled{3} which show a similar trend. 
The fact that  $\left<\Delta W_\text{neq}\right> < \left<\Delta U_\text{neq}\right> $ for small values of $t_\text{neq}$ immediately shows that heat is being converted into work. 
Only above $t_\text{neq}\approx 3s$, we recover the behavior of memory-free dissipative systems, i.e. that the amount of extracted work is smaller than the change of internal energy. 
The dashed curves correspond to the numerical results of our bath-particle model. 
In particular at small $t_\text{neq}$ we find excellent agreement, with some deviations (attributed to the simplicity of the model) towards larger timescales.
        

\begin{figure}
    \centering
    \includegraphics[width=8.6cm]{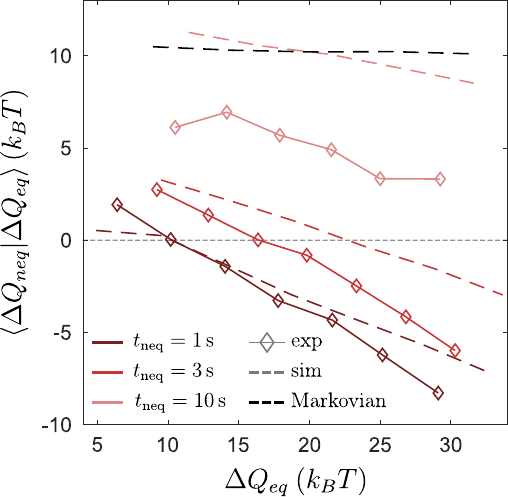}
    \caption{Correlation between $\Delta Q_\text{eq}$ and $\Delta Q_\text{neq}$ for $t_\text{neq}$ = 1, 3, and $\SI{10}{\s}$ (dark to light red). Symbols correspond to experimental data, dashed-lines to simulations. The black dashed-line corresponds to the Markovian case, which doesn't exhibit any correlation.
    }
    \label{fig:F4}
\end{figure}

Finally, we also expect a direct correlation between the heat dissipated in \circled{1} and measured in \circled{3}.
Due to energy conservation, such correlations should hold both for averages and single trajectories. 
Figure~\ref{fig:F4} shows the correlation of the distributions of $\Delta Q_\text{eq}$ and $\Delta Q_\text{neq}$ for $t_\text{neq}$ = 1, 3, and $\SI{10}{\s}$, respectively. 
Symbols and dashed lines correspond to experimental data and those obtained from simulations. 
The negative slope confirms that an increase of $\Delta Q_\text{eq}$ leads to a decrease of $\Delta Q_\text{neq}$ and thus to a larger ANH. 
As expected, with increasing $t_\text{neq}$, this correlation becomes weaker and eventually vanishes. 
We also performed Langevin simulations for a viscous fluid (black dotted line, see SM). 
Clearly, such heat correlations must be absent in memory-free viscous baths, but are a distinctive feature of non-Markovian surroundings.

Formally, the occurrence of an ANH is a consequence of the definitions of heat and work (Eqs.~\ref{dw eq} and~\ref{dq eq}) which have been derived for ideal heat baths with infinitely fast relaxation. 
In case of a non-Markovian heat bath, the slow decaying degrees of freedom must be taken into account in the definition of the above quantities. 
When explicitly considering the contribution of the bath particle and the spring in our model, a classical Markovian behavior is recovered and the ANH disappears. 
However, an accurate microscopic description is typically not achievable in complex systems. 
Therefore, the presence of an ANH in experiments can serve as a quantifier to provide evidence for (hidden) slow degrees of freedom in the bath.

\begin{figure}
    \centering
    \includegraphics[width=8.6cm]{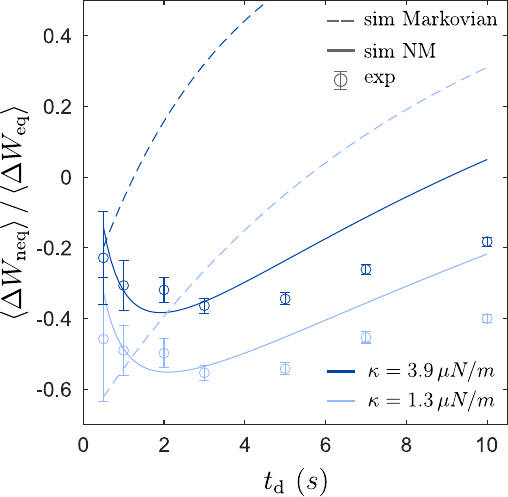}
    \caption{Cycle ratio $\frac{\left < \Delta W_\text{neq} \right > }{\left < \Delta W_\text{eq} \right > }$ as a function of  $t_\text{d}$, for  $\dot \gamma=\SI{0.2}{\micro\m\per\s}$ and $t_\text{neq}=\SI{1}{\s}$, with $\kappa=\SI{1.35}{\micro\N\per\m}$ (light blue) and $\kappa=\SI{3.89}{\micro\N\per\m}$ (dark blue). Symbols and plain lines correspond to non-Markovian experiments and simulations respectively. The dashed-lines correspond to simulations of a particle with friction $\gamma_{\rm N}=\gamma+\gamma_{\rm b}$ inside a Newtonian (Markovian) fluid.    
    }
    \label{fig:F5}
\end{figure}

In addition, the presence of ANH also has consequences for the design of energy efficient driving protocols in non-Markovian environments. 
Inspired by microscopic engines~\cite{martinez2016brownian,argun2017experimental,guevara2023brownian}, we have studied how the ratio $\frac{\left < \Delta W_\text{neq} \right > }{\left < \Delta W_\text{eq} \right > }$ varies as a function of the driving time $t_d$. 
Fig.5 shows the results obtained for $\dot\gamma=\SI{0.2}{\micro\per\s}$ and $t_\text{neq}=\SI{1}{\s}$, with $\kappa=\SI{1.35}{\micro\N\per\m}$ (light blue) and $\kappa=\SI{3.89}{\micro\N\per\m}$ (dark blue).
For the Newtonian case (dashed lines), we only show the results of our model (with $\gamma_{\rm N}=\gamma+\gamma_{\rm b}$ and $\kappa_\text{b}=0$).
As expected, in this situation $\frac{\Delta W_\text{neq}}{\Delta W_\text{eq}}$ monotonically increases with $t_\text{d}$: the longer the particle is driven, the more heat is dissipated in the bath, which decreases the fraction of work that can be recovered. 
Opposed to this, the behavior in a non-Markovian bath is quite different. 
Since heat can now be exploited to extract work, maximum efficiency is no longer achieved by avoiding heat generation, but by optimizing the amount of energy recovered.
As a result, the ratio $\frac{\Delta W_\text{neq}}{\Delta W_\text{eq}}$ now increases non-monotonically with $t_\text{d}$, and shows a minimum for $t_\text{d} \sim \SI{3}{\s}$. 
Notably, for larger values of $t_\text{d}$, and particularly in the case of the stiffer trap (dark blue), the measured efficiency in the non-Markovian bath is systematically better than in the Markovian one.
Simulations (plain lines) again nicely match the experimental data (symbols), with a very good agreement for small values of $t_\text{d}$, and some deviations afterward.


In summary, we have observed the onset of an averaged negative heat flow between a viscoelastic bath and a driven colloidal particle. 
Opposed to Markovian baths, where negative heat events only arise sporadically due to thermal fluctuations, our observations hold even when averaging over a large ensemble of trajectories. 
Our results, which are in good agreement with a micro-mechanical model, demonstrate that energy can be temporarily stored and recovered in baths with slow hidden degrees of freedom. 
As a consequence, non-Markovian baths can be exploited to increase the efficiency of cyclic processes compared to Markovian surroundings. 
We expect such approach to be important for the realization and optimization of novel types of microscopic heat engines. 
Finally, we our results do not only hold in case of viscoelastic baths but should also apply to any non-Markovian systems with a delayed response. 

\begin{acknowledgments}
We thank Samuel Monter for fruitful discussions.
This project was funded by the Deutsche Forschungsgemeinschaft (DFG), Grant No. SFB 1432 - Project ID 425217212.
\end{acknowledgments}



%

\end{document}